\documentclass[12pt,preprint]{aastex}
\begin{document}
\title{How the Coherent Tides Obstruct the Radial Infalls of Satellite Galaxies onto Clusters}
\author{Jounghun Lee}
\affil{Astronomy Program, Department of Physics and Astronomy, Seoul National University, 
Seoul 08826, Republic of Korea} 
\email{jounghun@astro.snu.ac.kr}
\begin{abstract}
A direct numerical evidence for the obstructing effect of the coherent tides on the infall-zone satellites around the cluster halos  
is presented. Analyzing the numerical data from a high-resolution N-body simulation, we calculate the mean fractions of the 
radial and tangential velocities of the infall-zone satellites around the cluster halos and investigate if and how they 
depend on the tidal coherence defined as the alignments between the major principal axes of the local tidal fields 
smoothed on the linear and nonlinear scales. It is found that the infall-zone satellites located in the regions with higher tidal coherence 
have significantly smaller and larger mean fractions of the radial and tangential velocities, respectively, which indicates that the radial 
infall of satellites onto host clusters are obstructed by the coherent tides.
We also show that those satellites separated by shorter distances from the host clusters, having lower-masses, and embedded in the 
anisotropic large-scale environments like filaments and sheets, are more vulnerable to the obstructing effect of the coherent tides. 
\end{abstract}
\keywords{cosmology:theory --- large-scale structure of universe}

\section{Introduction}\label{sec:intro}

A recent numerical study of \citet{zomg1} based on a zoomed N-body simulation has revealed that the thickness of a filament affects the 
growth history of its constituent halos.  According to its result, the galactic halos located in the conjunction of multiple fine filaments grow 
through "accretion" of matter onto them along the filaments, while the growths of those embedded in a broad bulky filament are quenched 
by the recession of matter from them in the plane perpendicular to the filament axis \citep[see also][]{zomg3}. Given this result, 
\citet{zomg1} claimed that the difference in the clustering pattern between the galactic halos embedded in the former and latter environments 
should be the origin of the halo assembly bias \citep{GW07}, which had been a long standing puzzle in the standard theory of 
structure formation. 

\citet{lee18a} suggested that it should be desirable to use a more quantitative unequivocal criterion for the classification of the tidal 
environments rather than using such a qualitative equivocal description as a bulky broad filament or multiple fine filaments for a systematic 
study of the large-scale tidal effect on the growth history of the halos.  Assuming that a broad bulky filament forms through two 
dimensional collapse of an over dense region along the coherent direction of the principal axes of the underlying tidal field over large 
scales, they introduced a new concept, "tidal coherence", quantitatively defining it as the alignments between the major principal axes of 
the local tidal fields smoothed on the nonlinear and linear scales.
Utilizing the data from the Sloan Digital Sky Survey Data Release 10 \citep{sdssdr10}, \citet{lee18a} detected a strong 
tendency for the less (more) luminous galaxies to populate the regions with higher (lower) tidal coherence even when the dominant 
effect of the local densities was nullified. \citet{lee18a} espoused the same logic of \citet{zomg1} to explain this observed tendency: The 
coherent tides quench the growth of the galaxies by obstructing the radial flow of satellites and matter onto the galaxies. 

In this Letter, we attempt to find a direct numerical evidence for the obstructing effect of the coherent tides on the satellites 
by examining the directions of the satellite motions relative to their hosts and their links to the degree of the tidal coherence in 
the regions where the hosts reside.  If the observed anti-correlation between the galaxy luminosity and the degree of the tidal 
coherence is truly caused by the obstructing effect of the coherent tides on the motions of satellites and matter as speculated by 
\citet{lee18a}, then those satellites located in the regions with higher tidal coherence must have relatively larger (smaller) fractions of the 
tangential (radial) velocities. To avoid poor number statistics and contaminations caused by low particle resolution, we will use as our host 
objects the cluster halos which have much larger number of satellites than the galactic halos. Throughout this Letter, we will assume a Planck 
$\Lambda$CDM ($\Lambda$ + cold dark matter) universe \citep{planck13}.  

\section{Numerical Analysis}\label{sec:analysis}

To investigate how the coherent tides affect the motions of the neighbor halos around the cluster halos, we analyze the numerical data 
extracted from the high-resolution Small MultiDark PLanck simulation (SMDPL) \citep{smdpl}, a pure DM $N$-body simulation performed as 
one  of the MultiDark simulation series \citep{multidark}.  The box size ($L_{\rm box}$), total number of DM particles ($N_{\rm p}$), and mass 
of an individual DM particle ($m_{\rm p}$) of the SMDPL are as follows: $L_{\rm box}=400\,h^{-1}$Mpc, $N_{\rm p}=3840^{3}$ and 
$m_{p}=9.63\times 10^{7}\,h^{-1}\,M_{\odot}$ (doi:10.17876/cosmosim/smdpl/).
\citet{smdpl} applied the Rockstar halo finder  \citep{rockstar} to the particle distributions from the SMPDL to compile a catalog of the 
DM halos and subhalos at various redshifts from $z=120$ to $z=0$. They also employed the cloud-in-cell method to construct the 
density field, $\delta({\bf x})$, of DM particles on a total of $512^{3}$ cubic grids which constitute the simulation volume at $z=0$ .

To measure the tidal coherence, $q$, at each of the $512^{3}$ grids, we use the tidal fields constructed from the density fields in our 
previous work \citep{lee18b} with the help of the Fast Fourier Transformation (FFT) code \citep{num_recipe}. 
Let us briefly describe how the construction of ${\bf T}({\bf x})$ proceeds:  
The first step is to smooth the density field with a Gaussian filter on a linear scale of $R_{f}=30\,h^{-1}$Mpc  and then to find the Fourier 
amplitude, $\tilde{\delta}({\bf x})$, of the smoothed density field, $\delta({\bf x})$,  via the FFT code. Here, the choice of the linear scale 
is made given the finding of \citet{PS17} that the influence of the large-scale environments on the galaxy properties persists at 
non-negligible level up to the scale of $30\,h^{-1}$Mpc. 

The second step is to compute the nine components of the Fourier amplitude of the tidal field as 
$\tilde{T}_{ij}=k_{i}k_{j}\tilde{\delta}({\bf k})/k^{2}$ where  $i,\ j \in \{1,2,3\}$ and ${\bf k}=(k_{i})$ is a wave vector dual to ${\bf x}$ in the 
Fourier space. The third step is to obtain the tidal field, ${\bf T}({\bf x})$, from $\tilde{\bf T}({\bf k})$ through the inverse FFT code and then 
to find a set of three eigenvalues, $\{\lambda_{1},\ \lambda_{2},\ \lambda_{3}\vert \lambda_{1}\ge\lambda_{2}\ge\lambda_{3}\}$ and 
corresponding eigenvectors, $\{{\bf e}_{1},\ {\bf e}_{2},\ {\bf e}_{3}\}$, via the similarity transformation of ${\bf T}({\bf x})$ at each grid. 
The whole process is repeated but on a nonlinear scale of $R^{\prime}_{f}=2\,h^{-1}$Mpc (a typical cluster size) to obtain a new set 
of three eigenvalues and eigenvectors, $\{\lambda^{\prime}_{1},\lambda^{\prime}_{2},\lambda^{\prime}_{3}\vert
\lambda^{\prime}_{1}\ge\lambda^{\prime}_{2}\ge\lambda^{\prime}_{3}\}$ and 
$\{{\bf e}^{\prime}_{1},\ {\bf e}^{\prime}_{2},\ {\bf e}^{\prime}_{3}\}$ at each grid.  

Following the original definition of \citet{lee18a}, we determine the tidal coherence, $q$, at each grid as 
$q = \vert{\bf e}_{1}\cdot {\bf e}^{\prime}_{1}\vert$ whose value lies in the range of $[0, 1]$. A larger value of $q$ translates into a 
highly coherent tide.  Here, a highly coherent tide represents the gravitational collapse of matter along the major principal axis 
of the local tidal field whose direction is coherent over a wide range of the scale (from $2\,h^{-1}$Mpc to $30\,h^{-1}$Mpc). 
From the Rockstar halo catalog at $z=0$ extracted from the CosmoSim database (doi:10.17876/cosmosim/smdpl/), 
we first select a total of $1659$ cluster-size halos with virial masses of $M_{h}\ge10^{14}\,h^{-1}M_{\odot}$. Tagging each of the selected 
cluster halos with the value of $q$ at the grid point where it is placed, we separate the cluster halos into four $q$-selected samples 
(see Table \ref{tab:sample}).  Hereafter, we are going to use interchangeably the terms, host cluster halos, host clusters, and hosts.

For each host, we search for its neighbor halos with virial masses of $M_{s}\ge 10^{11}\,h^{-1}M_{\odot}$ in the distance range of 
$1\le r/r_{v}<3$ with $r\equiv \vert{\bf r}\vert$, where $r_{v}$ is the virial radius of a host and ${\bf r}$ is the separation vector 
from the  host center. The DM halos with virial masses lower than $10^{11}\,h^{-1}M_{\odot}$ composed of less than $1000$ DM 
particles are excluded from the current analysis to avoid any numerical contamination caused by the low resolution. 
The distance range of $1\le r/r_{v}<3$ is called the infall zone where the gravitational influence of a host 
overwhelms the Hubble flow \citep{zu-etal14}. Expecting that the neighbor halos located in the infall zone of a host will become its 
satellites, we are going to refer to them as {\it infall-zone satellites} throughout this work. 

The fraction of the radial and tangential components of ${\bf v}$ are defined as 
$\delta v_{r}=\vert{\bf v}\cdot\hat{\bf r}\vert/v$ and $\delta v_{t}=\vert{\bf v}-{\bf v}_{r}\hat{\bf r}\vert/v$, respectively, 
where $\hat{\bf r}\equiv {\bf r}/r$ and $v\equiv \vert{\bf v}\vert$ (see Figure \ref{fig:config}). 
Finally, two ensemble averages, $\langle \delta v_{r}\rangle$ and $\langle \delta v_{t}\rangle$, are taken over the infall-zone satellites 
of the host clusters belonging to each $q$-selected sample. The errors in the measurement of the two averages, 
$\sigma_{t}$ and $\sigma_{r}$, are also estimated as the standard deviations in the mean values as 
$\sigma_{t}=\sqrt{\left[\langle(\delta v_{t})^{2}\rangle - \langle\delta v_{t}\rangle^{2}\right]/(N_{h}-1)}$ and 
$\sigma_{r}=\sqrt{\left[\langle(\delta v_{r})^{2}\rangle - \langle\delta v_{r}\rangle^{2}\right]/(N_{h}-1)}$ where $N_{h}$ is 
the number of the host halos belonging to each sample. 

Figure \ref{fig:tan_ec} shows the values of $\langle\delta v_{t}\rangle$ and $\langle\delta v_{r}\rangle$ with errors from the four $q$-
selected samples in the top and bottom panels, respectively. It can be clearly seen that the samples with higher-$q$ values yield higher 
and lower mean fractions of the tangential and radial velocities, respectively, indicating a detection of significant signals of 
$\langle\delta v_{t}\rangle$-$q$ correlation and $\langle\delta v_{r}\rangle$-$q$ anti-correlation. 
Although all of the four $q$-selected samples yield $\langle\delta v_{r}\rangle>\langle\delta v_{t}\rangle$, the difference 
between the two averages becomes smaller as $q$ increases. In fact, in the highest-$q$ range of $0.9<q\le1$ 
the value of $\langle\delta v_{r}\rangle$ becomes quite comparable to that of $\langle\delta v_{t}\rangle$. 

Yet, before interpreting the $\langle\delta v_{t}\rangle$-$q$ correlation and $\langle\delta v_{r}\rangle$-$q$ anti-correlation shown in Figure 
\ref{fig:tan_ec} as a direct evidence for the existence of the obstructing effect of the coherent tides on the radial motions of the infall-zone 
satellites,  an additional verification process is required: Since the same trend could be produced even when the coherent tides only facilitate 
the tangential motions of the infall-zone satellites without significantly obstructing their radial motions, it has to be investigated if and how the 
mean values of the tangential and radial velocities themselves vary with $q$. 
For this investigation, it is first necessary to deal with the differences among the four $q$-selected samples in the distributions of the 
host halo masses and separation distances of the infall-zone satellites, since the mean value of the total velocity, $\langle v\rangle$,  
depends most sensitively on $M_{h}$ and $r$. 
The top panel of Figure \ref{fig:mhdis} shows how the mass distributions of the host clusters from the four $q$-selected samples differ 
from one another.  Taking the same number of the host clusters at each bin of $M_{h}$, we control the four $q$-selected samples to have 
the identical mass distributions, which are shown in the bottom panel of Figure \ref{fig:mhdis}. In a similar manner, 
we also control further the four samples to have identical distributions of the separation distances of the infall-zone satellites around the host 
clusters. Figure \ref{fig:satdis} plots the original and controlled number distributions of the infall-zone satellites versus $r$ in the top and 
bottom panels, respectively.  

We compute the values of $\langle v_{t}\rangle$ and $\langle v_{r}\rangle$ from the four controlled samples and plot them 
versus $q$ in Figure \ref{fig:tan_msync_ec}, which clearly reveals the existence of both of the $\langle v_{t}\rangle$-$q$ correlation and 
$\langle v_{r}\rangle$-$q$ anti-correlation. Note, however,  that the $\langle v_{r}\rangle$-$q$ anti-correlation is more significant in strength 
than the $\langle v_{t}\rangle$-$q$ correlation: the value of $\langle v_{r}\rangle$ shows almost monotonic decrease with $q$ in the 
entire range of $q$, while that of $\langle v_{t}\rangle$ does not significantly increase with $q$ in the range of $0< q\le 0.7$, 
showing a rapid rise only in the highest $q$ range of $0.9<q\le 1$.  This result indicates that although the coherent tides simultaneously 
obstruct and facilitate the radial and tangential motions of the infall-zone satellites, respectively, the former effect is more dominant. It 
confirms the speculation of \citet{lee18a} that  the growths of the host clusters located in the regions with higher tidal coherence would be 
delayed compared with those in the regions with lower tidal coherence because the radial infalls of their satellites are obstructed by 
the coherent tides.

To see whether or not the degree of the vulnerability to the obstructing effect of the coherent tides depends on the masses 
of the infall-zone satellites, we take the averages,  $\langle\delta v_{t}\rangle$ and $\langle\delta v_{r}\rangle$, over the infall-zone satellites 
in the low-mass ($1\le M_{s}/[10^{11}\,h^{-1}\,M_{\odot}]<5$) and high-mass $(M_{s}/[10^{11}\,h^{-1}\,M_{\odot}]\ge 5)$ ranges separately,  
the results of which are shown in the left and right panels of Figure \ref{fig:tan_ec_mass}. The results shown in Figure \ref{fig:tan_ec} 
(i.e., the results obtained without constraining the range of $M_{s}$) are also shown as dotted lines for comparison in Figure 
\ref{fig:tan_ec_mass}. 
Both of the low-mass and high-mass cases exhibit the same trend as that shown in Figure \ref{fig:tan_ec}:  the $\langle\delta v_{t}\rangle$-$q$ 
correlation and the $\langle\delta v_{r}\rangle$-$q$ anti-correlation.  As can be witnessed, however, the low-mass infall-zone satellites yield 
much larger values of  $\langle\delta v_{t}\rangle$ than the high-mass counterparts in the whole range of $q$. 

Note that the value of $\langle\delta v_{t}\rangle$ exceeds that of $\langle\delta v_{r}\rangle$ in the range of $0.9\le q<1$ for 
the low-mass case while the increment of $\langle\delta v_{t}\rangle$ with $q$ or equivalently the decrement of 
$\langle\delta v_{r}\rangle$ with $q$ are more rapid for the high-mass case. 
Our explanation for this result is as follows. 
The strong gravitational attraction between the high-mass satellites and their host clusters can resist the obstructing effect of the 
weakly coherent tides (i.e., low value of $q$), while the low-mass satellites are so vulnerable to the tidal effects that their 
radial infall toward their hosts are readily obstructed even by the weakly coherent tides. 

Taking the averages over the satellites in the inner ($1\le r/r_{v}<2$) and outer ($2\le r/r_{v} <3$) infall zones separately, 
we investigate how the strengths of $\langle\delta v_{t}\rangle$-$q$ correlation and $\langle\delta v_{r}\rangle$-$q$ anti-correlation 
depend on the separation distance, $r$, the results of which are shown in Figure \ref{fig:tan_ec_dis}. 
As can be seen, the inner infall-zone satellites yield much larger values of $\langle\delta v_{t}\rangle$ than the outer infall-zone 
counterparts in the entire range of $q$. For the case of the inner infall-zone satellites, the value of $\langle\delta v_{t}\rangle$ is 
comparable to that $\langle\delta v_{r}\rangle$ even in the lowest $q$-range of $0\le q<0.4$. 
Furthermore, $\langle\delta v_{t}\rangle$ exceeds $\langle\delta v_{r}\rangle$ in the higher-$q$ range of $0.4\le q<1$. 
Whereas, the outer infall-zone satellites are found to have $\langle\delta v_{t}\rangle<\langle\delta v_{r}\rangle$ in the 
entire range of $q$, although they still exhibit the $\langle\delta v_{t}\rangle$-$q$ correlation and $\langle\delta v_{r}\rangle$-$q$ 
anti-correlation. This phenomenon can be understood given that the inner infall-zone satellites must have been exposed to the 
obstructing effect of the coherent tides for longer period of time than the outer infall-zone counterparts since they entered the 
infall zones of their hosts at earlier epochs. 

Another interesting issue is if and how the type of the cosmic web affects the behaviors of $\langle\delta v_{r}\rangle$ and 
$\langle\delta v_{t}\rangle$. To address this issue, we apply the conventional web-classification algorithm suggested by 
\citet{hah-etal07} to the eigenvalues, $\{\lambda_{1},\ \lambda_{2},\ \lambda_{3}\}$, of the tidal fields smoothed on the linear scale 
of $R_{f}=30\,h^{-1}$Mpc and then determine the web type of a grid at which each host resides. If all of the tidal eigenvalues, 
$\lambda_{1},\ \lambda_{2},\ \lambda_{3}$, at a given grid are positive (negative), then the web-type of the grid is determined to be knot 
(void).  If only the lowest (highest) eigenvalue is negative (positive), then the web-type is filament (sheet).  The member clusters belonging 
to each sample are tagged with the classified web-types of the grids in which their positions lie. 

Breaking each of the four $q$-selected samples into four web type selected subsamples, we redo the whole analysis with the infall-zone 
satellites around the host clusters belonging to each subsample, the results of which are depicted in Figures 
\ref{fig:tan_ec_web1}-\ref{fig:tan_ec_web2}.  As can be seen, the same trend of the $\langle\delta v_{t}\rangle$-$q$ correlation and 
$\langle\delta v_{r}\rangle$-$q$ anti-correlation is found in the filament and sheet environments.  As can be seen, in the highest-$q$ 
range of $0.9<q\le 1$, the values of $\langle\delta v_{t}\rangle$ exceed those of $\langle\delta v_{r}\rangle$ for both of the cases 
of the filaments and sheets. But, note also that while the values of $\langle\delta v_{t}\rangle$ from the filaments are larger than those 
from the sheets in the low-$q$ range of $0<q\le 0.7$, almost no difference is found in the values of $\langle\delta v_{t}\rangle$ between 
the filaments and the sheets in the high-$q$ range of $0.7<q\le 1$. 

The results shown in Figure \ref{fig:tan_ec_web1} contain two crucial implications. First, obstruction of the radial motions of the infall-zone 
satellites by the weakly coherent tides ($0<q\le 0.7$) is the most effective in the filament environments, which is consistent with the claim of 
\citet{zomg1}. Second, obstruction of the radial flow of the satellites by the strongly coherent tides ($0.7<q\le 1$) in the sheets can be 
as effective as in the filaments.  For the case of a filament,  its thickness was used as an indicator of the tidal coherence 
\citep[e.g.,][]{zomg1}. For the case of a sheet, however, we suggest that the tidal coherence should be best reflected by its {\it flatness}, 
predicting that the growths of the hosts embedded in large-scale flat sheets would be delayed by the obstructing 
effect of the coherent tides on their satellites.  

Meanwhile, the knot and void environments yield somewhat different behaviors of $\langle\delta v_{t}\rangle$ and 
$\langle\delta v_{r}\rangle$ against the variation of $q$ (Figure \ref{fig:tan_ec_web2}). In the range of $0<q\le 0.9$, the two 
mean fractions exhibit the same trends: monotonic increment and decrement  of $\langle\delta v_{t}\rangle$ and 
$\langle\delta v_{r}\rangle$ with the increment of $q$, respectively.  
However, in the highest-$q$ range of $0.9<q\le 1$, abrupt down-turn of $\langle\delta v_{t}\rangle$ and up-turn of 
$\langle\delta v_{r}\rangle$ are witnessed.  Note in particular that the void environments yield the lowest value of 
$\langle\delta v_{t}\rangle$ in the highest-$q$ range of $0.9\le q< 1$ rather than in the lowest-$q$ range, although the errors are large. 
This result indicates that the coherent tides in the isotropic environments like knots and voids can facilitate the radial motions of 
the infall-zone satellites rather than obstruct them. 
It may be explained as follows. In the isotropic environment where ${\bf e}_{1}$ and ${\bf e}^{\prime}_{1}$ are almost perfectly aligned 
with each other (i.e., $0.9<q\le 1$), the direction of ${\bf e}_{3}$ is also likely to be strongly aligned with ${\bf e}^{\prime}_{3}$. 
Since in the isotropic knot (void) environments, the gravitational compression (rarefaction) of matter occur 
along all of the three principal axes simultaneously, the strong ${\bf e}_{3}$-${\bf e}^{\prime}_{3}$ alignment will have an effect of 
facilitating the radial infall of the satellite galaxies onto the host clusters.

Given the results shown in Figures \ref{fig:tan_ec_web1}-\ref{fig:tan_ec_web2} that the anisotropic environments like filaments and sheets 
produce stronger obstructing effect of the tidal coherence, it is suspected that at a given region the tidal coherence $q$ should be 
correlated with the tidal strength, which is often quantified in terms of the ellipticity, $\epsilon$, of the local potential field
as $\epsilon=(\lambda_{1}-\lambda_{3})/(3+\delta)$ with $\delta=\sum_{i=1}^{3}\lambda_{i}$ \citep{yan-etal13}. 
To examine whether or not  the detected signal of $\langle\delta v_{t}\rangle$-$q$ correlation and $\langle\delta v_{r}\rangle$-$q$
anti-correlation is at least partly caused by the differences in tidal strengths among the four $q$-selected samples,  we measure the tidal 
strengths, $\epsilon$, at the locations of the host clusters and determine the $\epsilon$ distribution of the host clusters belonging to  
each sample. 

As can be seen in the top panel of Figure \ref{fig:elldis}, the four $q$-selected samples yield different distributions of $N_{h}({\epsilon})$, 
which indicates that the detected signal of  $\langle\delta v_{t}\rangle$-$q$ correlation and $\langle\delta v_{r}\rangle$-$q$
anti-correlation should not be completely independent of the $q$-$\epsilon$ correlation.  To nullify the effect of the tidal strengths, we create 
four resamples of the host clusters according to the values of $q$, controlling them to have the identical distributions of $N_{h}(\epsilon)$, which 
are shown in the bottom panel of Figure \ref{fig:elldis}. Then, we recalculate $\langle\delta v_{t}\rangle$ and $\langle\delta v_{r}\rangle$ from 
the four controlled resamples, which are plotted in Figure \ref{fig:tan_ec_sync}. As can be seen, the resamples still show 
clear signals of the $\langle\delta v_{t}\rangle$-$q$ correlation and $\langle\delta v_{r}\rangle$-$q$ anti-correlation, although the strengths 
of the signals are slightly reduced compared with the original ones (dotted lines).  
We emphasize here that since the four controlled resamples have the identical $\epsilon$-distributions, they have no difference in the tidal 
strengths. If the observed signals of the $\langle\delta v_{t}\rangle$-$q$ correlation and $\langle\delta v_{r}\rangle$-$q$ anti-correlation from 
the original samples were completely due to the differences in the tidal strengths among the samples, then the four controlled 
resamples would yield zero signals. This result confirms the existence of a {\it dominant} obstructing effect of the tidal coherence $q$ on the 
cluster satellites, independent of the $q$-$\epsilon$ correlation.

\section{Summary and Discussion}\label{sec:dis}

\begin{figure}
\begin{center}
\plotone{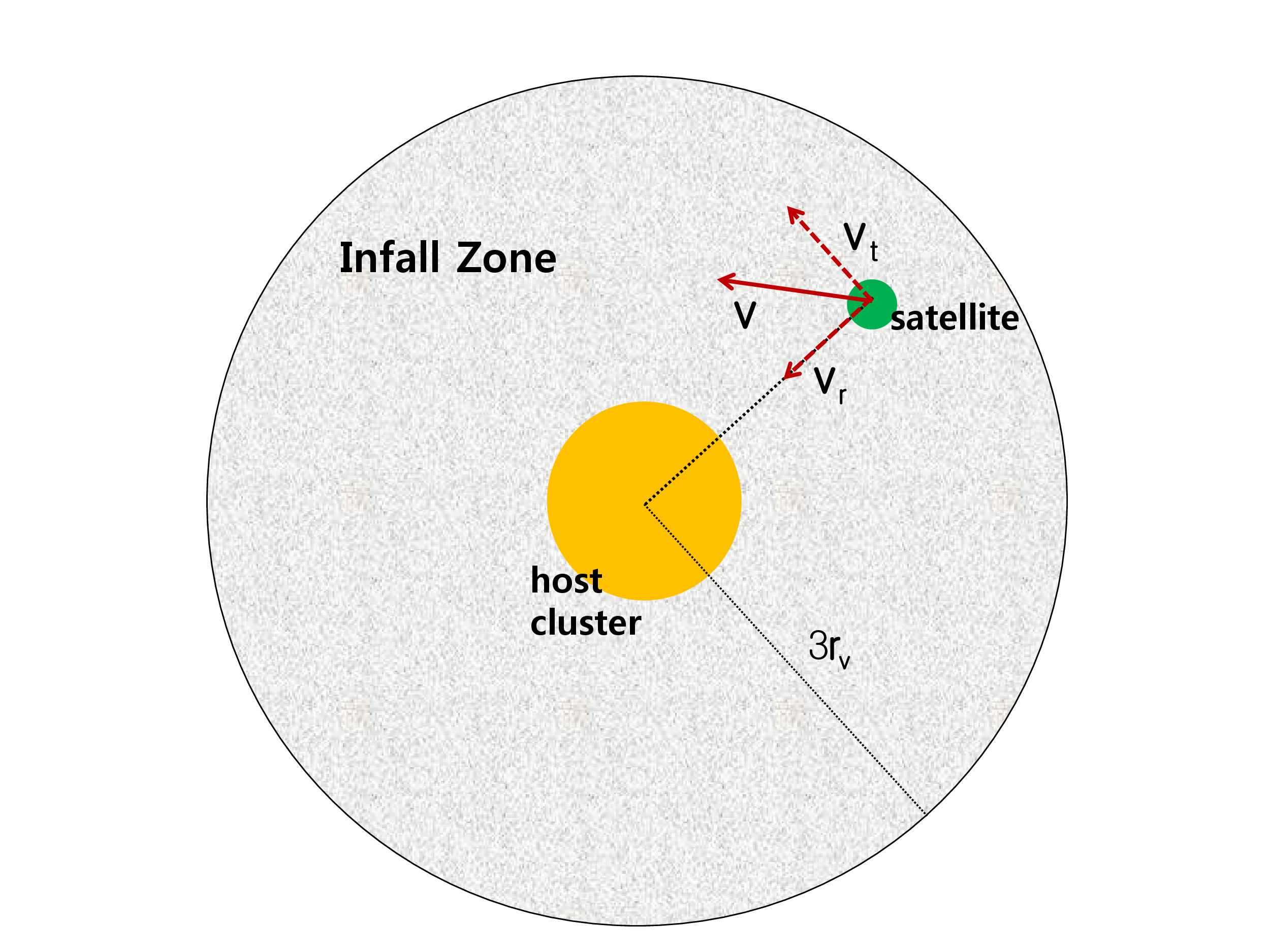}
\caption{Illustration of an infall-zone satellite whose comoving peculiar velocity ${\bf v}$ can be decomposed into a radial (${\bf v}_{r}$) 
and a tangential $({\bf v}_{t}$) component in the infall zone around a host cluster with virial radius $r_{v}$.}
\label{fig:config}
\end{center}
\end{figure}

\begin{figure}
\begin{center}
\plotone{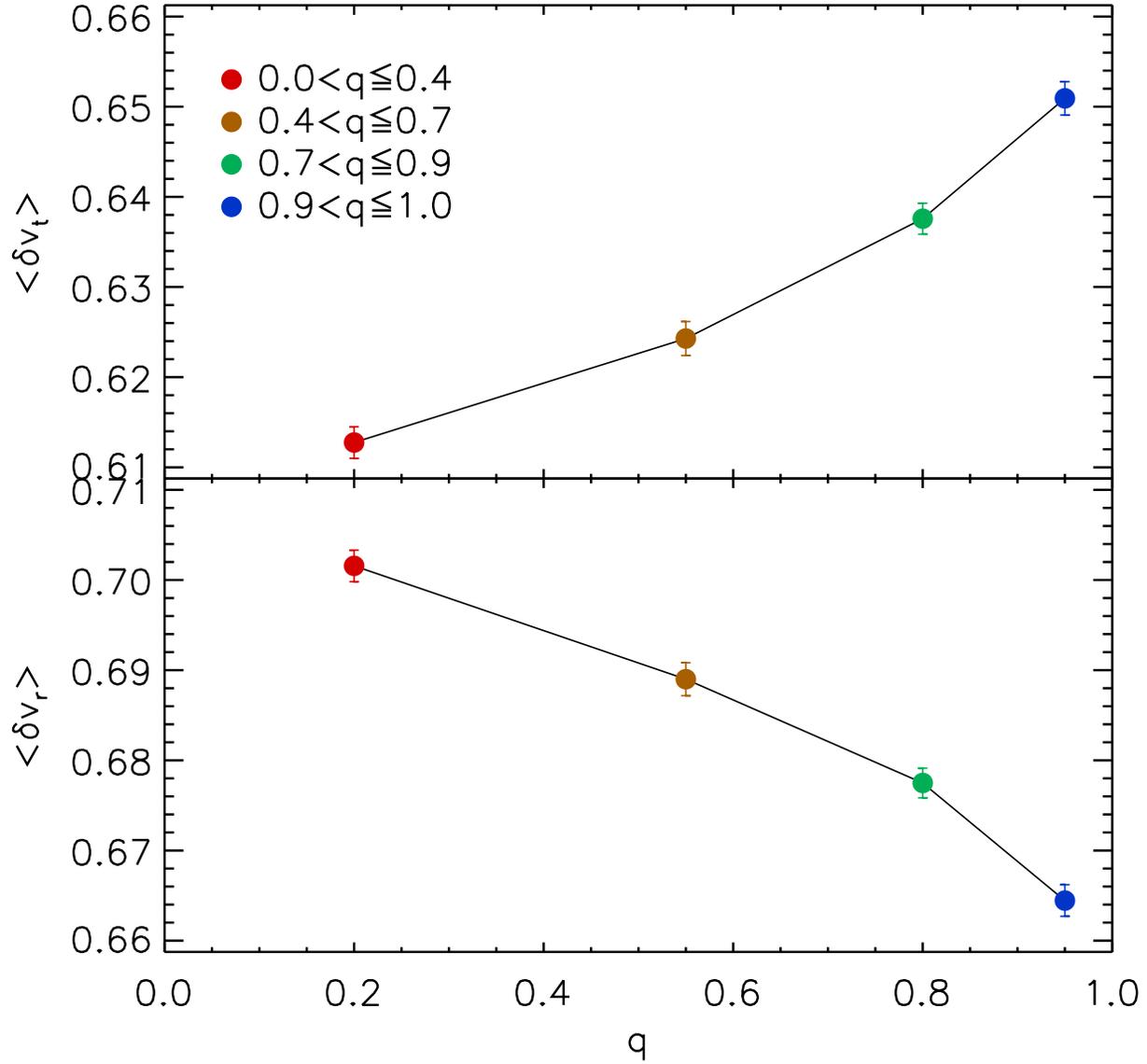}
\caption{Mean fractions of the tangential and radial velocities of the infall-zone satellites versus 
the tidal coherence, $q$, in the top and bottom panels, respectively.}
\label{fig:tan_ec}
\end{center}
\end{figure}

Analyzing the Rockstar halo catalog and density field at $z=0$ from the SMDPL \citep{smdpl}, we have located the infall-zone satellites 
($M_{s}\ge 10^{11}\,h^{-1}M_{\odot}$) around the host clusters ($M_{h}\ge 10^{14}\,h^{-1}M_{\odot}$) in the distance range of 
$1\le r/r_{v}<3$, and then determined two directions of the major principal axes of the tidal field smoothed on the scales of $2\,h^{-1}$Mpc 
and $30\,h^{-1}$Mpc at the position of each host. Calculating the tidal coherence, $q$, as the alignments between the two directions of the 
major principal axes for each cluster, we have created four $q$-selected samples of the host clusters in the ranges of 
$0<q\le 0.4$, $0.4< q\le 0.7$, $0.7<q\le 0.9$ and $0.9<q\le 1$.  
For each sample, we have measured the tangential and radial components of the relative velocities of the infall-zone satellites around 
the hosts and taken the mean values (Figure \ref{fig:config}). It has turned out that a sample with a higher $q$ value yields significantly 
higher and lower mean fractions of the tangential and radial velocities of the infall-zone satellites, respectively (Figure \ref{fig:tan_ec}). 

\begin{figure}
\begin{center}
\plotone{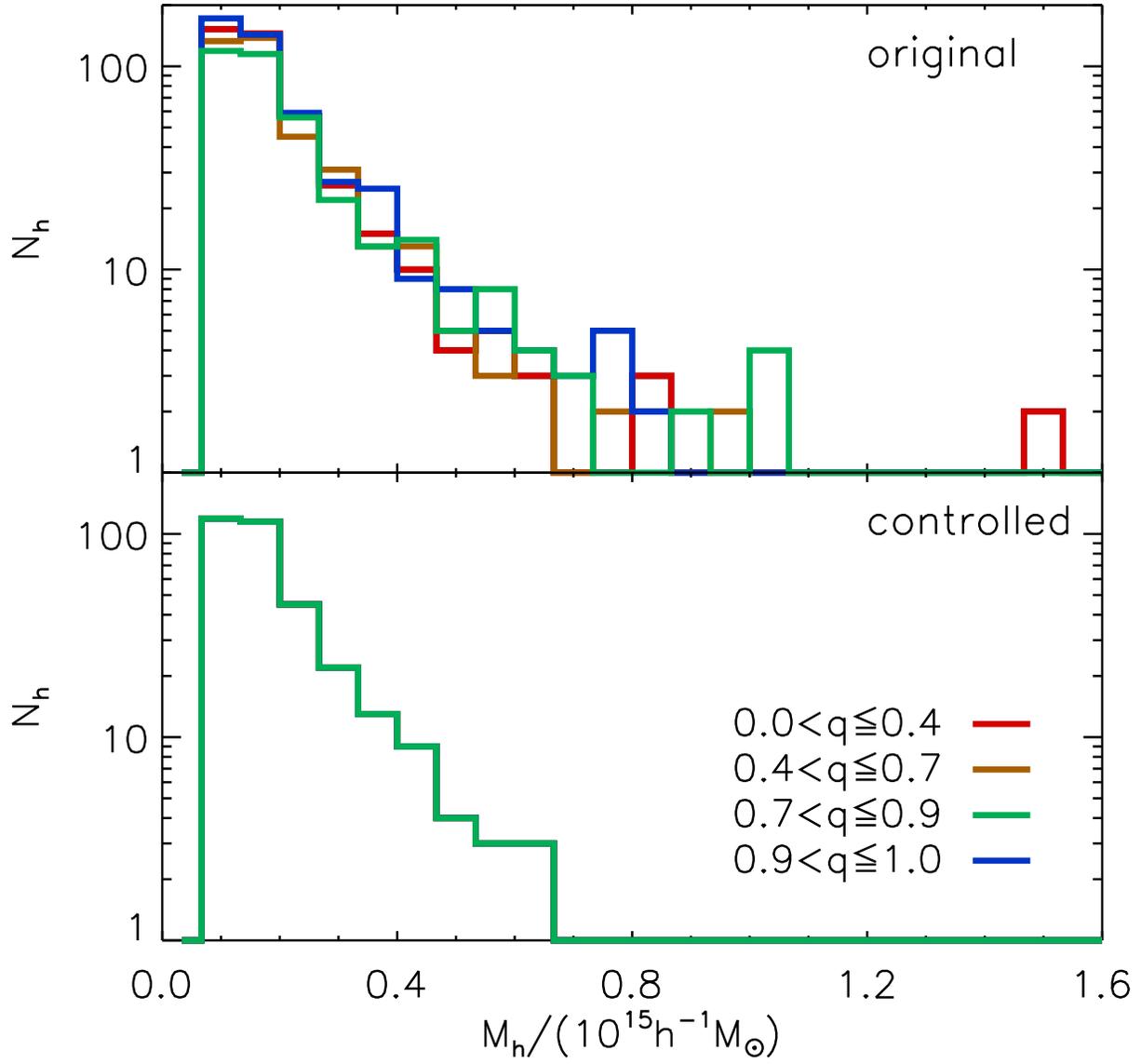}
\caption{Original (top panel) and controlled (bottom panel) number distributions of the clusters as a 
function of their virial masses from four $q$-selected samples.} 
\label{fig:mhdis}
\end{center}
\end{figure}

\begin{figure}
\begin{center}
\plotone{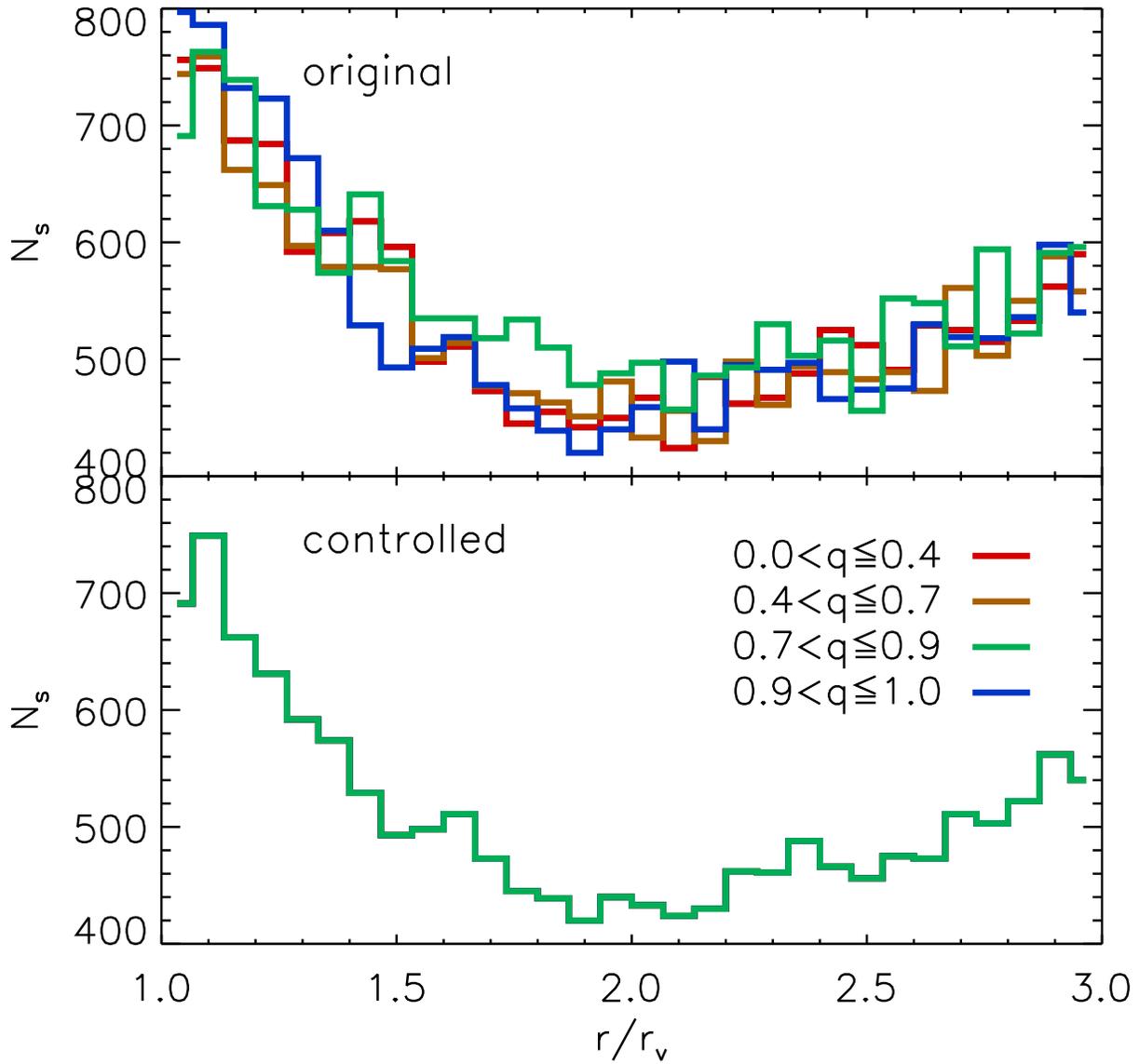}
\caption{Original (top panel) and controlled (bottom panel) number distributions of the infall-zone satellites 
as a function of their separation distances from the centers of their hosts from the four $q$-selected samples.} 
\label{fig:satdis}
\end{center}
\end{figure}

\begin{figure}
\begin{center}
\plotone{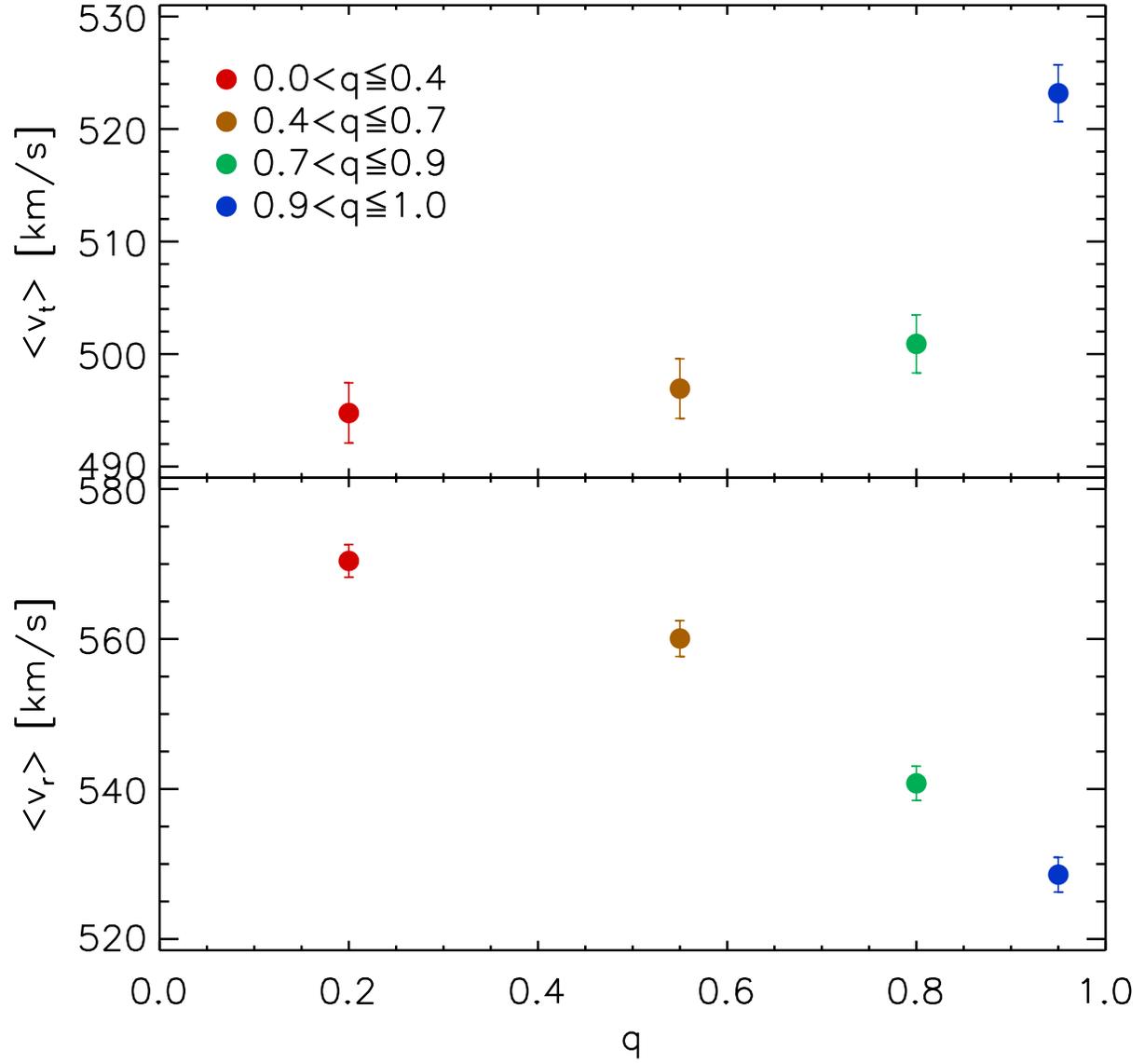}
\caption{Mean tangential and radial velocities of the infall-zone satellites from the 
controlled resamples versus $q$ in the top and bottom panels, respectively.} 
\label{fig:tan_msync_ec}
\end{center}
\end{figure}

 Creating the four $q$-selected resamples of the host clusters which are controlled to have identical distributions of $M_{h}$ and $r$ 
 (Figures \ref{fig:mhdis}-\ref{fig:satdis}), we have computed the mean values of the tangential and radial velocities from each of the controlled 
 samples. It has been found that the mean radial velocity decreases monotonically with $q$ in the entire range of $q$ while the mean 
tangential velocity shows slight increase with $q$ in the range of $0< q\le 0.7$ and a rapid rise in the range of $0.9< q\le 1$. 
Our interpretation of the results shown in Figures \ref{fig:tan_ec}-\ref{fig:tan_msync_ec} is that the gravitational compression of matter 
along the major principal axis whose direction is coherent over widely separated scales (i.e., the coherent tide) has an effect of significantly 
obstructing the radial motions of the infall-zone satellites toward the host clusters, while it tends to facilitate their tangential motions.  

\begin{figure}
\begin{center}
\plotone{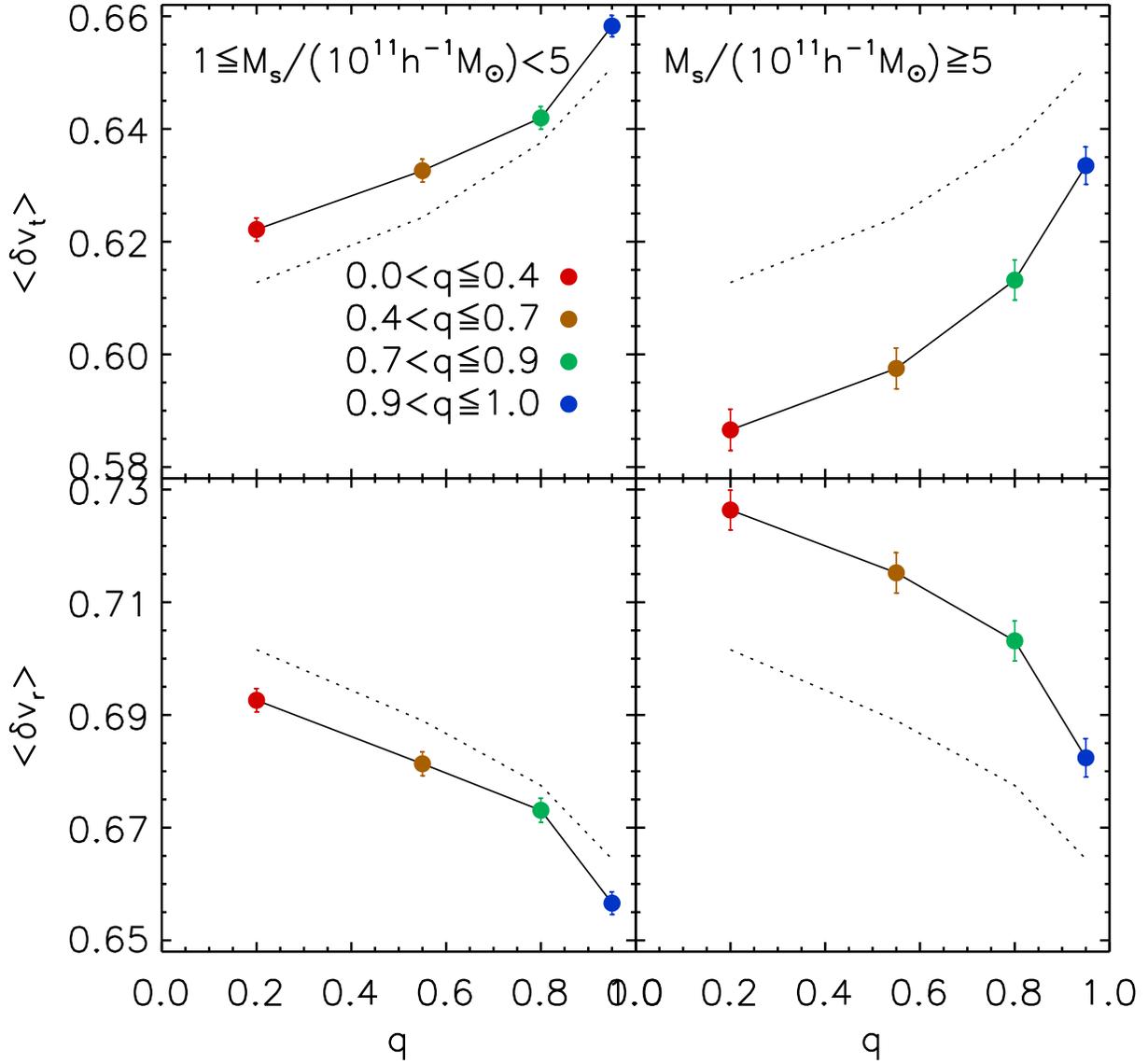}
\caption{Same as Figure \ref{fig:tan_ec} but averaged over the low-mass (left panel) and high-mass 
(right panel) infall-zone satellites separately. The results shown in the top and bottom panels of 
Figure \ref{fig:tan_ec} are also shown as dotted lines here for comparison.}
\label{fig:tan_ec_mass}
\end{center}
\end{figure}

\begin{figure}
\begin{center}
\plotone{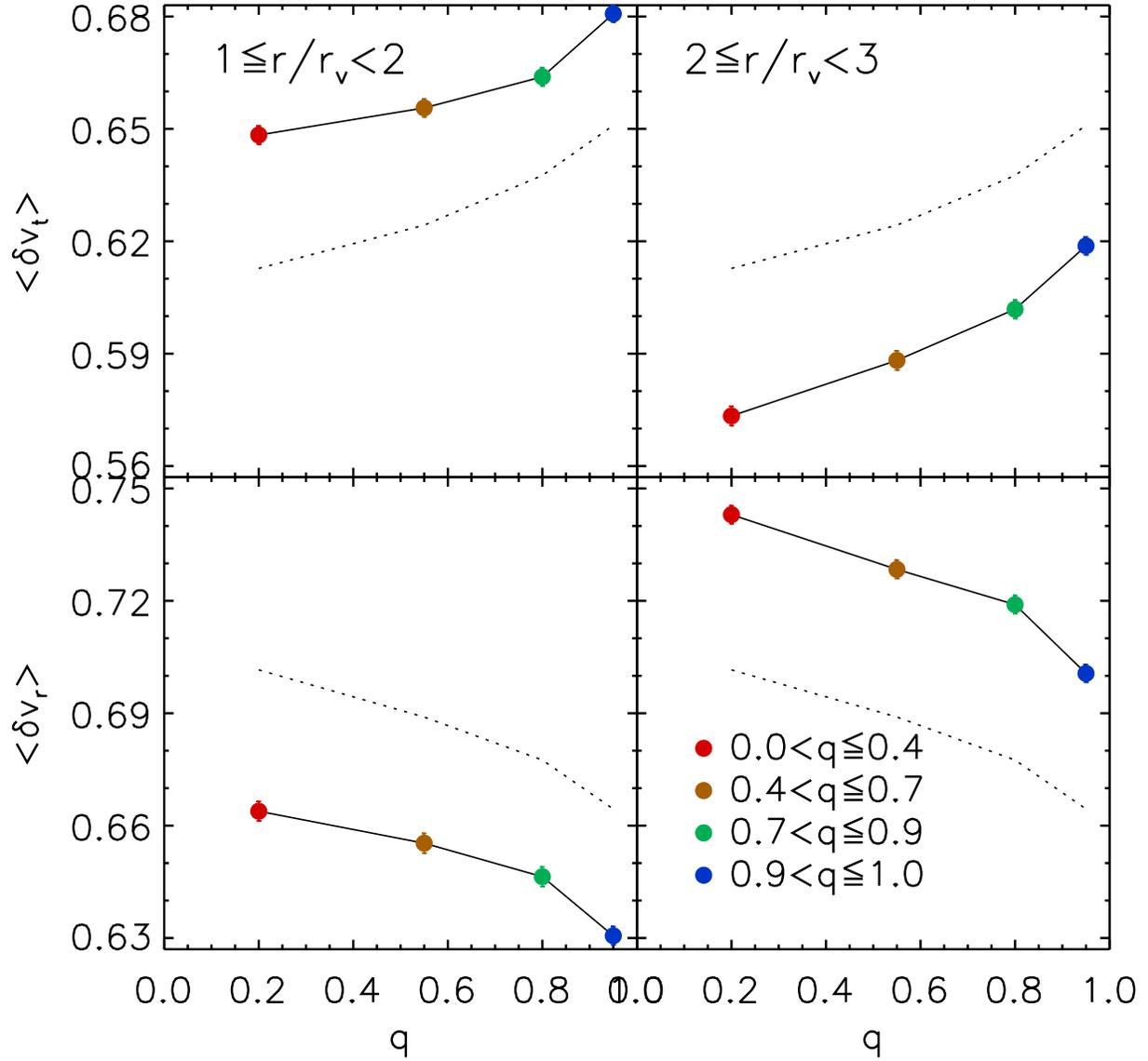}
\caption{Same as Figure \ref{fig:tan_ec} but averaged over the inner (left panel) and outer  
(right panel) infall-zone satellites separately.}
\label{fig:tan_ec_dis}
\end{center}
\end{figure}

\begin{figure}
\begin{center}
\plotone{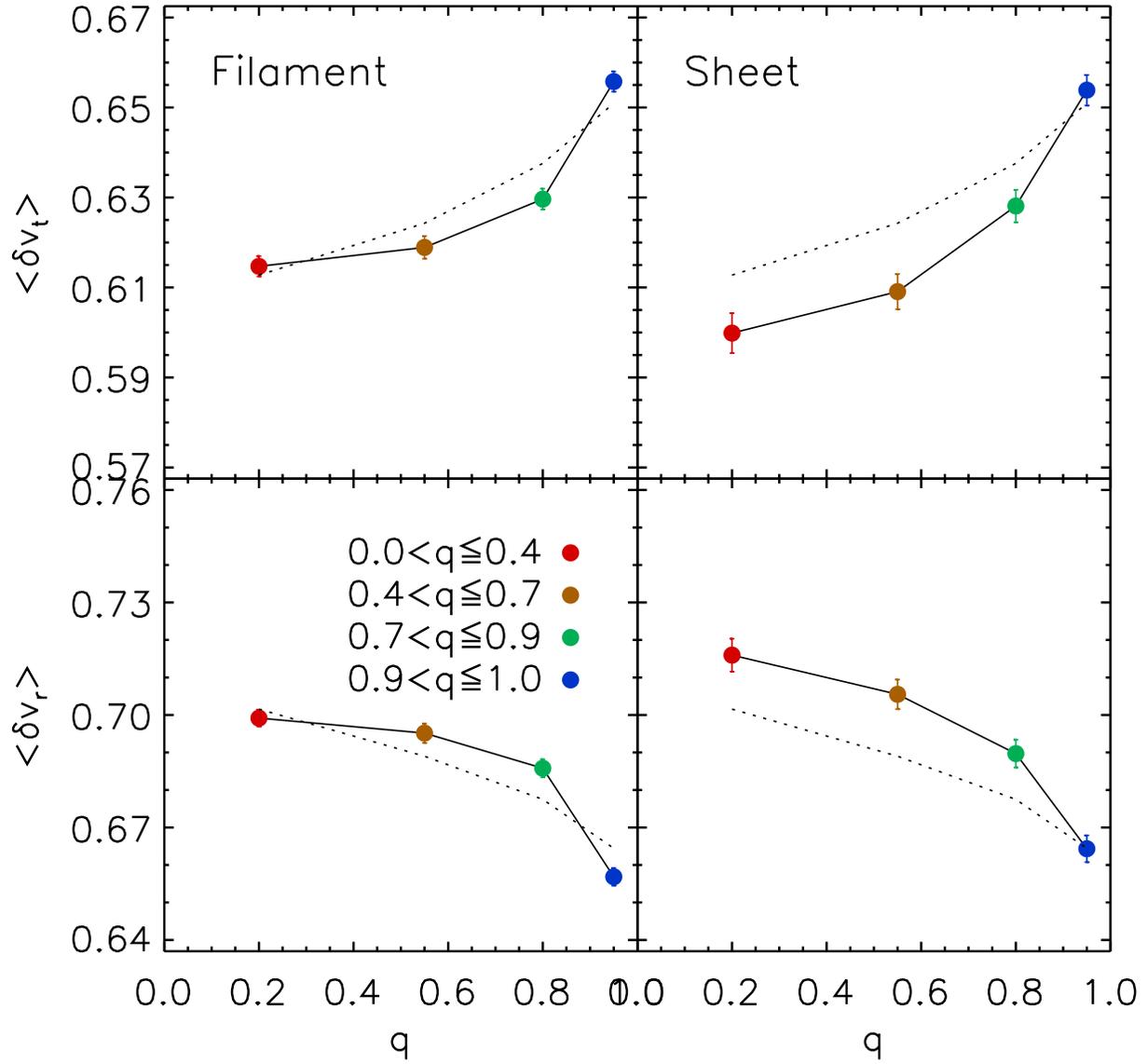}
\caption{Same as Figure \ref{fig:tan_ec} but averaged over the infall-zone satellites around the clusters 
embedded in the filaments (left panel) and sheets (right panel) separately.}
\label{fig:tan_ec_web1}
\end{center}
\end{figure}

\begin{figure}
\begin{center}
\plotone{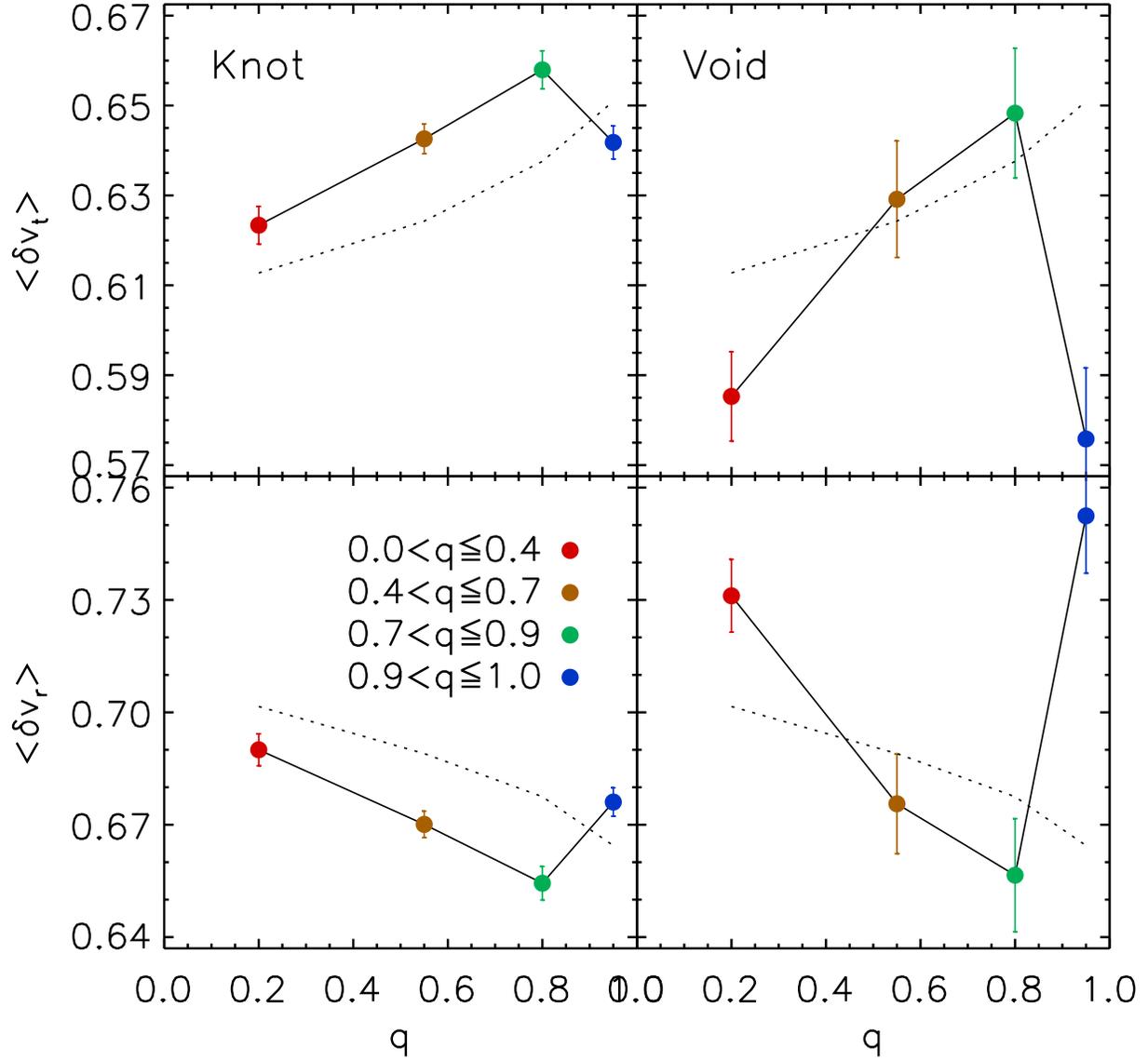}
\caption{Same as Figure \ref{fig:tan_ec} but averaged over the infall-zone satellites around the clusters 
embedded in the knots (left panel) and voids (right panel) separately.}
\label{fig:tan_ec_web2}
\end{center}
\end{figure}

We have also investigated if and how the obstructing effect of the coherent tides depends on the satellite mass, separation distance and 
web-type of the large-scale environment. It has been shown that the infall-zone satellites having lower masses 
($M_{s}\le 5\times 10^{11}\,h^{-1}M_{\odot}$), located in the inner infall zone ($1\le r/r_{v}< 2$), and surrounded by the anisotropic 
environments like filaments and sheets are more vulnerable to the obstructing effect of the coherent tides than the other cases 
(Figures \ref{fig:tan_ec_mass}-\ref{fig:tan_ec_web1}). 
No matter what distances and masses the infall-zone satellites have, their mean fractions of the tangential velocities have been found to 
monotonically increase with the increment of $q$. However, in the isotropic large-scale environments like knots and voids, the mean 
fractions of the tangential velocities exhibit abrupt decrease rather than increase in the highest-$q$ range, $0.9<q\le 1$ 
(Figure \ref{fig:tan_ec_web2}).  To explain this phenomenon, we have noted that the almost perfect alignments of the major 
principal axes of the tidal fields between the linear and nonlinear scales (i.e., $0.9<q\le 1$) would induce the strong alignments 
of the minor principal axes. Since in the isotropic environments the gravitational compression of matter occur concurrently along all of the 
three principal axes of the tidal fields, the radial infall of the satellite galaxies onto the host clusters will be facilitated rather than obstructed 
by the gravitational compression of surrounding matter distribution along the coherent directions of the minor principal axes in the isotropic 
environments with highest tidal coherence.

In this work, we have implicitly assumed that the web-classification algorithm of \citet{hah-etal07} can be used to determine the web 
types of environments even when the tidal fields are smoothed on the linear scale of $R_{f}=30\,h^{-1}$Mpc. However, it has to be mentioned 
here that this assumption has yet to be justified  since the previous works which tested this algorithm usually considered much smaller scales, 
$R_{f}\le 5\,h^{-1}$Mpc \citep[e.g., see][]{hah-etal07,trace_web18}.  It will be necessary to examine up to what scale 
this algorithm is applicable for the determination of the web types, which is, however, beyond the scope of this Letter.

\begin{figure}
\begin{center}
\plotone{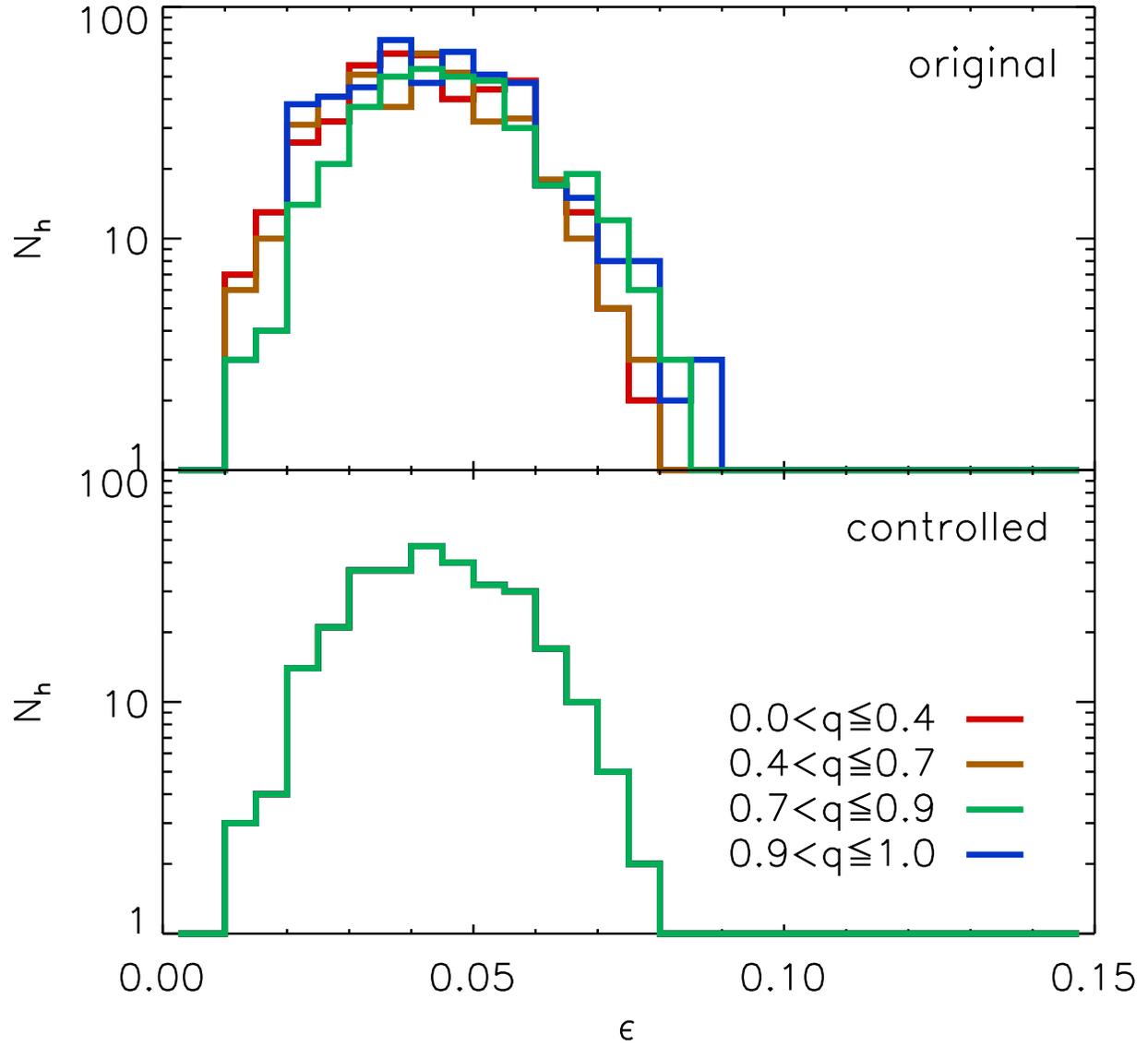}
\caption{Original (top panel) and controlled (bottom panel) number distributions of the clusters as a 
function of the ellipticities of the regions where the host clusters are located.} 
\label{fig:elldis}
\end{center}
\end{figure}

\begin{figure}
\begin{center}
\plotone{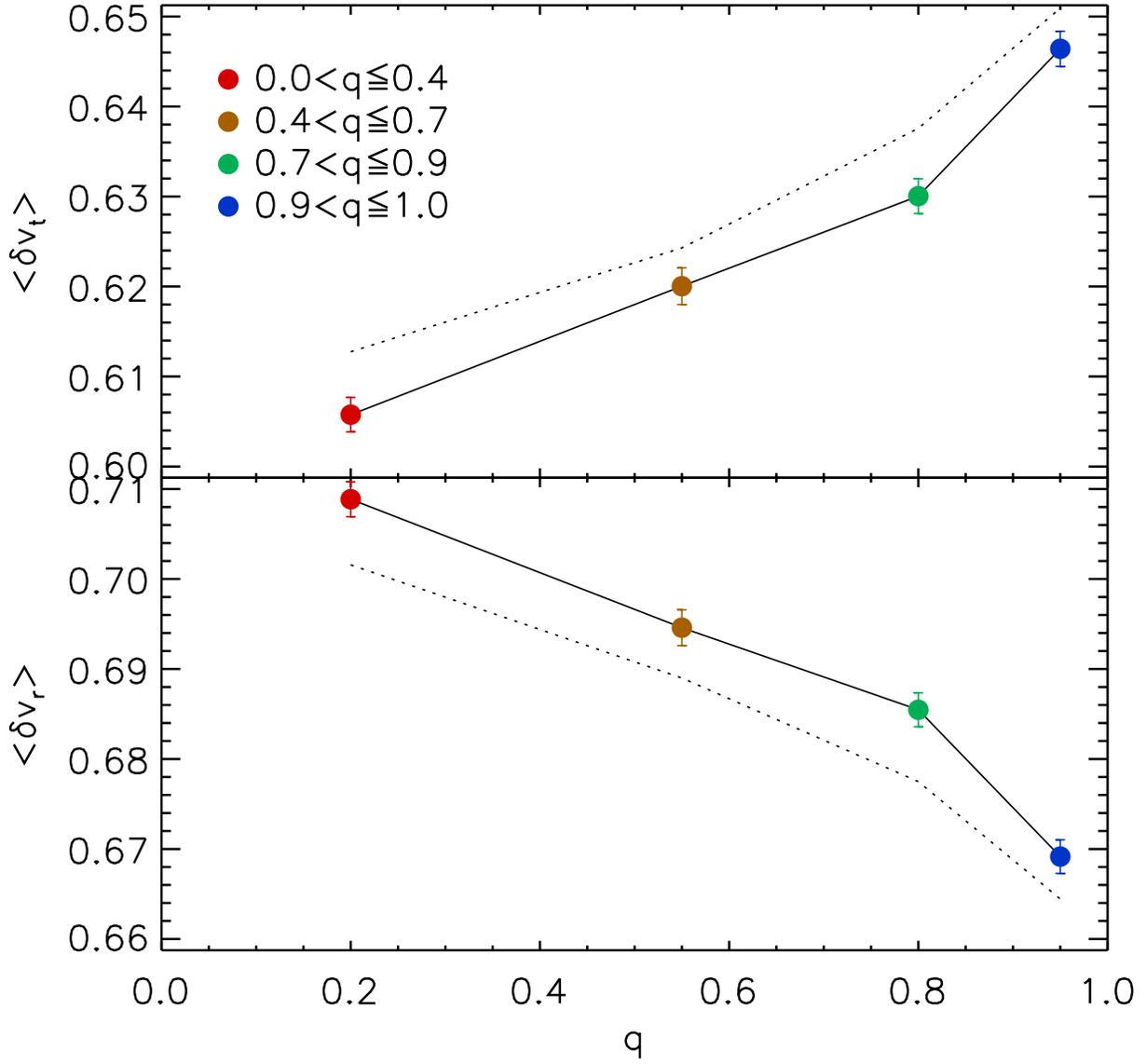}
\caption{Same as Figure \ref{fig:tan_ec} but from the controlled resamples of the host clusters.}
\label{fig:tan_ec_sync}
\end{center}
\end{figure}

We have also nullified the dependence of the obstructing effect of the tidal coherence on the tidal strengths by creating the four 
$q$-selected resamples of the host clusters which have identical distributions of the ellipticities of the potential field (a measure of the 
tidal strength). Showing that the four resamples still yield clear signals of the correlations (anti-correlations) between the mean fractions of the 
tangential (radial) velocities and the degree of the tidal coherence, we have confirmed that the obstructing effect of the tidal coherence on the 
cluster satellites is independent of and dominant over the effect of the tidal strengths (Figures \ref{fig:elldis}-\ref{fig:tan_ec_sync}).  

Although our observational evidence for the obstructing effect of the coherent tides has been found from the cluster satellites 
rather than from the galaxy ones, we believe that it can be used to explain the recently discovered anti-correlation between the tidal 
coherence and the galaxy luminosity from the SDSS DR10 \citep{lee18a}: The galaxies located in the regions with higher tidal coherence 
are less luminous, because the coherent tides facilitate the tangential motions of the satellites and obstruct  their radial flow onto the host 
galaxies, as conjectured by \citet{lee18a}.
Our study also provides a fundamental explanation for the phenomenon found by \citet{PB06} that the spatial distributions of the 
less luminous galaxies tend to be more anisotropic. The highly coherent tides often produce large-scale bulky filaments and flat sheets 
in which the spatial distributions of the galaxies are highly anisotropic. Thus, the observed link of the luminosities of the 
galaxies with the anisotropic shape of the large-scale structure where they reside is in fact a reflection of the strong 
obstructing effect of the coherent tides.   

As mentioned in Section \ref{sec:intro}, this work was motivated and inspired by the recent numerical study of \citet{zomg1} which 
revealed that the growth history of the galaxies depend on the thickness of the host filaments. Our findings supports their idea, and 
extends its validity to the cluster scales and to the other types of the cosmic web by showing that the obstruction of the radial motions 
of the cluster satellites by the coherent tides occur not only in the filaments but also in the sheets, knots and voids.  
Given the current results obtained from the cluster satellites at $z=0$, we expect that the lower-mass galaxy satellites must be much more 
vulnerable to the obstructing effect of the coherent tides than the cluster satellites. We also expect that the coherent tides must have a 
weaker obstructing effect at higher redshifts since the gravitational compression of matter along the major principal axes of the large-scale 
tidal fields at higher redshifts would be incomplete.  In the isotropic environments like knots and voids, however, the 
effect of the tidal coherence is expected to be stronger at higher redshifts than at $z=0$. Our future work is in the direction of testing these 
expectations against numerical data from N-body simulations with much higher resolution. 

\acknowledgements

ÒThe CosmoSim database used in this paper is a service by the Leibniz-Institute for Astrophysics Potsdam (AIP).
The MultiDark database was developed in cooperation with the Spanish MultiDark Consolider Project CSD2009-00064.Ó
I gratefully acknowledge the Gauss Centre for Supercomputing e.V. (www.gauss-centre.eu) and the Partnership for Advanced 
Supercomputing in Europe (PRACE, www.prace-ri.eu) for funding the MultiDark simulation project by providing computing time on 
the GCS Supercomputer SuperMUC at Leibniz Supercomputing Centre (LRZ, www.lrz.de).
The Bolshoi simulations have been performed within the Bolshoi project of the University of California High-Performance 
AstroComputing Center (UC-HiPACC) and were run at the NASA Ames Research Center.

I am grateful to an anonymous referee for very helpful comments which helped me correctly interpret the final result of the current work.
I acknowledge the support of the Basic Science Research Program through the National Research Foundation (NRF) of Korea 
funded by the Ministry of Education (NO. 2016R1D1A1A09918491).  I was also partially supported by a research grant from the 
NRF of Korea to the Center for Galaxy Evolution Research (No.2017R1A5A1070354).

\clearpage
\clearpage
\clearpage
\clearpage
\clearpage
\clearpage
\clearpage
\clearpage
\clearpage
\clearpage
\clearpage
\clearpage
\begin{deluxetable}{ccc}
\tablewidth{0pt}
\setlength{\tabcolsep}{5mm}
\tablecaption{Cluster numbers and mass ranges from four $q$-selected samples}
\tablehead{$q$ & $N_{h}$ & $M_{h}$ \\
& & $(10^{14}\,h^{-1}M_{\odot})$ }
\startdata
$(\,0.0, \   0.4\,]$  & $429$  &[1.0, 17.1] \\
$(\,0.4 , \   0.7\,]$  &$395$ & [1.0, 12.8] \\
$(\,0.7, \    0.9\,]$ &$466$ & [1.0, 14.9]  \\
$(\,0.9, \   1.0\,]$  & $369$ & [1.0, 14.4] 
\enddata
\label{tab:sample}
\end{deluxetable}


\clearpage
\begin{thebibliography}{000}
\bibitem[Ahn et al.(2014)]{sdssdr10} 
Ahn, C.~P., Alexandroff, R., Allende Prieto, C., et al.\ 2014, \apjs, 211, 17 
\bibitem[Behroozi et al.(2013)]{rockstar} 
Behroozi, P.~S., Wechsler, R.~H., \& Wu, H.-Y.\ 2013, \apj, 762, 109 
\bibitem[Borzyszkowski et al.(2017)]{zomg1} 
Borzyszkowski, M., Porciani, C., Romano-D{\'{\i}}az, E., \& Garaldi, E.\ 2017, \mnras, 469, 594
\bibitem[Gao \& White(2007)]{GW07} 
Gao, L., \& White, S.~D.~M. \ 2007, \mnras, 377, L5
\bibitem[Garaldi et al.(2017)]{zomg3} 
Garaldi, E., Romano-D{\'{\i}}az, E., Borzyszkowski, M., \& Porciani, C.\ 2017, arXiv:1707.01108 
\bibitem[Hahn et al.(2007)]{hah-etal07} 
Hahn, O., Porciani, C., Carollo, C.~M., \& Dekel, A.\ 2007, \mnras, 375, 489 
\bibitem[Klypin et al.(2016)]{smdpl} 
Klypin, A., Yepes, G., Gottl{\"o}ber, S., Prada, F., \& He{\ss}, S.\ 2016, \mnras, 457, 4340
\bibitem[Lee(2018a)]{lee18a} 
Lee, J.\ 2018a, \apj, 867, 36
\bibitem[Lee(2018b)]{lee18b} 
Lee, J.\ 2018b, arXiv:1808.08559
\bibitem[Libeskind et al.(2018)]{trace_web18} 
Libeskind, N.~I., van de Weygaert, R., Cautun, M., et al.\ 2018, \mnras, 473, 1195 
\bibitem[Pandey \& Bharadwaj(2006)]{PB06} 
Pandey, B., \& Bharadwaj, S.\ 2006, \mnras, 372, 827 
\bibitem[Pandey \& Sarkar(2017)]{PS17} 
Pandey, B., \& Sarkar, S.\ 2017, \mnras, 467, L6 
\bibitem[Press et al.(1992)]{num_recipe} 
Press, W.~H., Teukolsky, S.~A., Vetterling, W.~T., \& Flannery, B.~P.\ 1992, Cambridge: University Press, |c1992, 2nd ed.,
\bibitem[Planck Collaboration et al.(2014)]{planck13} 
Planck Collaboration, Ade, P.~A.~R., Aghanim, N., et al.\ 2014, \aap, 571, A16
\bibitem[Riebe et al.(2013)]{multidark} 
Riebe, K., Partl, A.~M., Enke, H., et al.\ 2013, Astronomische Nachrichten, 334, 691 
\bibitem[Yan et al.(2013)]{yan-etal13} 
Yan, H., Fan, Z., \& White, S.~D.~M.\ 2013, \mnras, 430, 3432
\bibitem[Zu et al.(2014)]{zu-etal14} 
Zu, Y., Weinberg, D.~H., Jennings, E., Li, B., \& Wyman, M.\ 2014, \mnras, 445, 1885
\end{thebibliography}
\end{document}